\documentclass[prd,aps,preprint,preprintnumbers,nofootinbib]{revtex4}
\usepackage{epsfig,footnote}
\usepackage{ulem}
\usepackage{color}
\usepackage{array}
\usepackage{amssymb}
\usepackage{amsmath}
\usepackage{graphicx,subfigure}
\usepackage{longtable}
\usepackage{verbatim}
\usepackage{amsfonts}
\usepackage{hyperref}
\usepackage{cancel}
\usepackage{adjustbox}
\usepackage{graphicx}

\begin{document}
\title{Updated branching ratios and $CP$ asymmetries in $D \to PV$ decays  }
\author{Hui Zheng, Jia-Rui Dong and Si-Hong Zhou\footnote{corresponding author: shzhou@imu.edu.cn}}
\affiliation{School of Physical Science and Technology, 
 Inner Mongolia Key Laboratory of 
Microscale Physics and Atomic Manufacturing, 
Inner Mongolia University, Hohhot 010021, China}
\affiliation{Center for Quantum Physics and Technologies, 
School of Physical Science and Technology, 
Inner Mongolia University, Hohhot 010021, China}

\begin{abstract}
Motivated by extensive new high-precision experimental data,
we present an updated analysis of the two-body charm decays
$D \to PV$ (with $P =\pi,K, \eta^{(\prime)}$ and
 $V =\rho, K^*, \omega, \phi$) within the 
 factorization-assisted topological-amplitude (FAT) approach.
In the framework, flavor SU(3) symmetry breaking effect
is incorporated into the topological amplitudes, allowing
the nonfactorizable contributions from topological diagrams 
to be expressed as a minimal set of universal parameters 
determined through a global fit to experimental data. 
Thanks to the sufficient data with high precision, we are 
now able to quantify the nonfactorizable contribution of 
so-called ``factorization" $T$ diagram, which is essential 
for explaining the observed branching ratios. The parameters 
for the $C$, $E$, and $A$ diagrams are also updated with 
significant improved precision, and notably, the resulting 
strong phases differ substantially from those in the earlier 
FAT analysis. We find that the $C$ topological amplitude 
features substantial nonfactorizable effects in the charm sector.
These refined parameters enable predictions of 
significantly improved accuracy, yielding branching ratios 
in good agreement with current data and the latest results 
in topological diagram approach, and the predicted direct 
$\mathit{CP}$ asymmetries differ distinctly form previous 
FAT results due to updated strong phases. In several modes, 
the $\mathit{CP}$ asymmetries are predicted to reach 
$\mathcal{O}(10^{-3})$, thereby making them promising 
observables for future high-precision experiments 
at  LHCb, BESIII and Belle II. Predictions for unobserved 
decay modes, especially those with branching fractions 
of order $10^{-4}\sim10^{-3}$, are also provided for 
forthcoming experimental tests.
\end{abstract}

\maketitle

\section{Introduction}\label{Introduction}
The charm quark, being neither heavy nor light with a mass 
of order 1.5 GeV, places $D$-meson decays at the interface 
of the perturbative and nonpertubative QCD. Consequently, 
they provide a unique testing ground for studying the 
corresponding decay mechanism. Furthermore, the study of  
$\mathit{CP}$ violation in $D$-meson can probe for new physics. 
Given that  $\mathit{CP}$ asymmetries in the Standard Model 
are predicted to be of order $10^{-4}$, any significant observation 
of $\mathit{CP}$ asymmetries in hadronic charm decays would be 
a direct indication of physics beyond the standard model.

The theoretical description of exclusive hadronic $D$ decay 
based on QCD remains an open challenge, primarily due to 
the intermediate scale $m_c$. This situation contrasts with 
that of nonleptonic $B$ decays, where QCD-inspired approach 
like QCD factorization, perturbative QCD, and soft-collinear 
effective theory are successfully applicable, as they rely on a
heavy quark expansion in power of $1/m_b$. For charm sector, 
the charm mass $m_c$, however, is not sufficiently heavy to 
allow for a sensible application of such an expansion.
As a consequence, several model-independent approaches 
based on symmetry have been proposed. 
A prominent example is the topological diagram 
approach~\cite{Cheng:2010ry, Bhattacharya:2009ps}, 
which classifies flavor-flow diagrams according to the topologies 
of weak interactions, incorporating all strong interaction effects. 
Instead of a complicated QCD analysis, this method parametrizes 
the corresponding topological diagram amplitudes, thereby 
relying on experimental measurements for the extraction of 
these parameters. In 2010, since a limited number of data 
points from Cabibbo-favored modes were available, only 
a subset of the topological diagram parameters could be 
extracted~\cite{Cheng:2010ry}. To estimate the penguin 
diagram contributions necessary for calculation of direct 
 $\mathit{CP}$ asymmetries, the authors subsequently 
 employed the QCD factorization approach within the 
 same topological framework~\cite{,Cheng:2012wr}. 
 For a systematic and global analysis of tree and penguin 
 contributions in $D$ decays, the FAT approach was later 
 developed~\cite{Li:2012cfa, Li:2013xsa}, which successfully 
 reproduced both the branching ratios and direct $\mathit{CP}$ 
 asymmetries, and notably resolved the long-standing puzzle 
 of the large difference between the $D^0 \to \pi^+ \pi^-$ and
$D^0 \to K^+ K^-$ branching ratios. In particular, the FAT 
prediction for $\Delta A_\mathit{CP}^\mathrm{SM}\equiv 
 A_\mathit{CP} (D^0 \to K^+ K^-)-A_\mathit{CP} (D^0 \to \pi^+ \pi^-)
 = (-0.57 \sim -1.87) \times 10^{-3}$ is consistent with the first 
 observation of $\mathit{CP}$ violation by the LHCb Collaboration 
 in the charm sector, $\Delta A_\mathit{CP}^\mathrm{SM}
=(-1.54\pm 0.29) \times 10^{-3}$~\cite{LHCb:2019hro}.

Since the 2010s, significant experimental progress has been 
made, with numerous new decay modes being measured 
and the precision of known modes continually improved,
including the first observation of $\mathit{CP}$ violation 
in the $D \to PP$ decay channels. Enabled by increasingly 
precise experimental data, the topological diagram approach 
has been continuously refined. This includes global analysis 
of $D \to VP$ and their $\mathit{CP}$ violation in 
2016~\cite{Cheng:2016ejf}, 2019~\cite{Cheng:2019ggx}, 
and 2021~\cite{Cheng:2021yrn}, respectively, as well as 
a comprehensive update of the branching ratios for all 
$D \to PP, PV, VV$ decay modes in 2024~\cite{Cheng:2024hdo}. 
These successive studies have yielded progressively
more accurate and precise predictions for the branching ratios 
of two-body $D$ decays. Given this progress and ongoing 
accumulation of experimental results, a corresponding 
update of the $D$-meson decays within the FAT approach 
has become imperative.

In this work, we perform a systematic analysis of 
$D \to PV$ decays using the FAT approaches, 
incorporating the latest experimental results.
The FAT framework was originally established for 
charm meson decays~\cite{Li:2012cfa, Li:2013xsa}
and subsequently generalized to $B$-meson decays 
by one of us (S.-H. Z.) and 
collaborators~\cite{ Zhou:2015jba, Zhou:2016jkv, Wang:2017hxe} 
 to systematically address nonfactorizable contributions 
 in heavy meson decays. The approach has since been 
 successfully applied to various phenomena, including 
 $D^0-\bar D^0$ mixing~\cite{Jiang:2017zwr}, 
 $K_S^0-K_L^0$ asymmetries~\cite{Wang:2017ksn}
and CP violation~\cite{Yu:2017oky} in charm decays 
into neutral kaons. It has also been used to extract the 
CKM phase $\gamma$ from charmless two-body $B$ 
decays~\cite{Zhou:2019crd}, while a comprehensive 
review of the methodology is available in~\cite{Qin:2021tve}.
More recently, the framework has been extended to both 
quasi-two-body $B$ 
decays~\cite{Zhou:2021yys,Zhou:2023lbc,Zhou:2024qmm,Ou-Yang:2025ije} 
and $D$ decays~\cite{Zhou:2025nao,Wang:2025rkr}. Building upon the 
topological diagram approach~\cite{Cheng:2010ry,Cheng:2012wr}, 
the FAT approach systematically categorizes decay amplitudes 
into distinct topological diagrams based on electroweak interactions,
while explicitly incorporating flavor SU(3) symmetry breaking effects.
Specifically, the symmetry breaking effects are implemented in the
topological diagram amplitudes mainly through the factorization 
of form factors and decay constants, assisted by QCD factorization. 
The remaining nonfactorizable contributions in the topological 
diagram amplitudes are characterized by a minimal set of 
universal parameters, which are constrained through a global 
fit to all available experimental data. By preserving SU(3)
symmetry breaking effects, the FAT approach enables a 
global fit using all Cabibbo-favored (CF), 
singly Cabibbo-suppressed (SCS), and doubly Cabibbo-suppressed 
(DCS) modes of $D \to PV$. In contrast, the topological diagram 
approach does not perform such a global analysis.
 Instead, it extracts the topological amplitudes exclusively 
 from CF modes~\cite{Cheng:2024hdo} and uses them to 
 predict SCS and DCS decays. To improve agreement 
 with experimental SCS results, the topological diagram 
 approach introduces SU(3) breaking corrections 
 between SCS and CF modes via a naive factorization 
 prescription. A further key advantage of the FAT approach 
 lies in its unified description of penguin and tree-level 
 contributions within a single coherent framework. This 
 differs fundamentally from the topological diagram approach,
where penguin amplitudes are typically estimated using 
QCD factorization, while tree-level topologies are treated 
diagrammatically. In the present FAT analysis, the inclusion 
of newer and more precise experimental data allow us to 
determine significantly refined nonfactorizable parameters 
compared to earlier FAT results~\cite{Li:2013xsa}.
These improvements translate directly into more accurate 
predictions for both branching ratios and direct $\mathit{CP}$ 
asymmetries in $D \to PV$ decays, with particularly notable 
differences appearing in $\mathit{CP}$ violation observables 
due to the updated determinations of strong phases.

 The remainder of this paper is organized as follows. 
 Section~\ref{sec:2} introduces the theoretical framework. 
Numerical results and detailed discussions arepresented in Sec.~\ref{sec:3}. 
Finally, we conclude in Sec.~\ref{sec:4}.


\section{Factorization Amplitudes for Topological Diagrams}\label{sec:2}
\begin{figure} [htb]
 \begin{center}
 \scalebox{1}{\epsfig{file=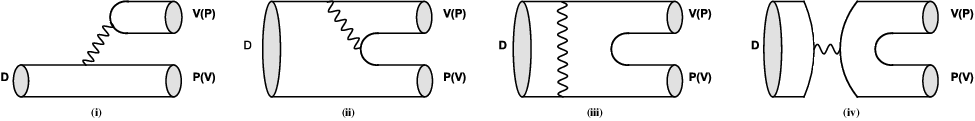}}
 \caption{Topological tree diagrams for $D \to P \, V$ 
 decay with a wavy line representing the $W$ boson:
   (i) color-favored tree diagram $T$,
   (ii) color-suppressed tree diagram $C$, 
   (iii) $W$-exchange tree diagram $E$,
   and (iv) $W$-annihilation tree diagram $A$.}
  \label{tree}
 \end{center}
  \end{figure}

\begin{figure} [htb]
 \begin{center}
 \scalebox{1}{\epsfig{file=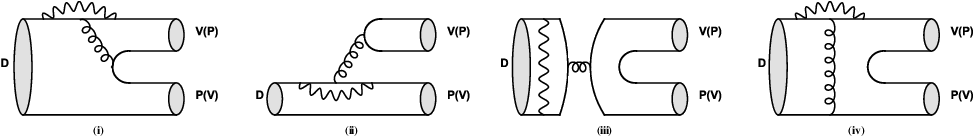}}
 \caption{Topological penguin diagrams for $D \to P \, V$ 
 decay with a wavy line representing the $W$ boson and 
a curly line the gluon:
   (i) color-favored penguin diagram $PT$,
   (ii) color-suppressed penguin diagram $PC$, 
   (iii) $W$-exchange penguin diagram $PE$,
   and (iv) $W$-annihilation penguin diagram $PA$.}
  \label{penguin}
  \vspace{-0.7cm}
 \end{center}
  \end{figure}

The two-body $D$-meson decays proceed through 
the quark flavor transitions by $c \to d(s)\,  u \bar d (\bar s) $
at leading order and $c \to u\,  q \bar q \, (q=u,\, d,\, s)$ at 
next-to-leading order in the electroweak interaction.
Correspondingly, the topological diagrams consist of 
tree diagrams and penguin diagrams, as shown in 
Figs.~\ref{tree} and Fig.~\ref{penguin}, respectively. 
Based on the topological structures of the weak 
interactions, the tree diagrams are conventionally 
classified into four types:
(a) Color-favored tree emission diagram $T$, 
(b) Color-suppressed tree emission diagram $C$,  
(c) $W$-exchange tree diagram $E$, and 
(d) $W$-annihilation tree diagram $A$.
The penguin topologies are categorized analogously,  
with ``tree" replaced by ``penguin", and are denoted 
as $PT$, $PC$, $PE$ and $PA$, mirroring the tree 
diagram types. Unlike charmless $B$ decays where 
penguin contributions dominate in certain channels 
such as $B \to K\, \pi$~\cite{Zhou:2016jkv}, the 
corresponding penguin amplitudes in charm sector 
are suppressed by both the Wilson coefficients and 
the Cabibbo-Kobayashi-Maskawa (CKM) matrix 
elements. We therefore neglect these penguin 
contributions in the our analysis of branching 
fractions while retaining them for $\mathit{CP}$ 
violation observables. 
 
 For the amplitudes corresponding to the aforementioned 
 topological diagrams, we adopt the parametrization scheme
 for $D \to PV$ decays as described in Ref.~\cite{Li:2013xsa} 
within the FAT framework. Specifically, nonfactorizable 
 contributions are parametrized by $\chi$ and $\phi$, 
 representing their magnitudes and associate strong phases, 
 respectively.  A key modification relative to Ref.~\cite{Li:2013xsa} 
 is the introduction of an additional parameter to account for 
 nonfactorizable contributions to ``factorization" $T$ diagram. 
 Compared to $B$ decays, $D$ processes 
 exhibit significantly larger nonfactorizable contributions,
even in the $T$ diagram, owing to their lower energy release.
Consequently, neglecting the nonfactorizable components 
of the $T$ diagram is no longer a suitable approximation 
in the high-precision studies of charm decays. 

The amplitudes for the various topological diagrams 
are given as follows, with the $T$ diagram amplitude 
incorporating nonfactorizable contributions,
\begin{align}\label{2amp}
\begin{aligned}
T_{PV}(C_{PV})&=\sqrt{2}\, G_{F}\, V_{cq}^*V_{uq^{'}}\, 
a_1(\mu)\, \left( a_2^V(\mu)\right) \, f_V \, m_V\,
F_1^{DP}(m_V^2)\, (\varepsilon^{*}_{V} \cdot p_D),
\\
T_{VP}(C_{VP})&= \sqrt{2}\, G_{F}\, V_{cq}^*V_{uq^{'}}\, 
a_1(\mu)\left(a_2^P(\mu)\right)f_{P} m_V
A_0^{DV}(m_{P}^2)\, (\varepsilon^{*}_{V} \cdot p_D),
\\
E_{PV,VP}&=\sqrt{2}\, G_{F}\, V_{cq}^*V_{uq^{'}}\,  
C_2(\mu)\, \chi^E_{q(s)}\, \mathrm{e}^{i\phi^E_{q(s)} }
f_D\, m_V\,  \frac{f_{P}\, f_V}{f_\pi \, f_\rho}\, (\varepsilon^{*}_{V} \cdot p_D),
\\
A_{PV, VP}&=\sqrt{2}\, G_{F}\, V_{cq}^*V_{uq^{'}}\, 
C_1(\mu)\, \chi^A_{q(s)}\, \mathrm{e}^{i\phi^A_{q(s)} }
f_D\, m_V\,  \frac{f_{P}\, f_V}{f_\pi \, f_\rho}\, (\varepsilon^{*}_{V} \cdot p_D)\, ,
\end{aligned}
\end{align}
with 
\begin{align}\label{a1a2}
\begin{aligned}
a_1(\mu)&=C_2(\mu)\, +\, C_1(\mu)\left[{1\over N_C}\, +\, \chi^T\right],\\
a_2^{V(P)}(\mu)&=C_1(\mu)\, +\, C_2(\mu)
\left[{1\over N_C}\, +\, \chi_{V(P)}^C \, \mathrm{e}^{i\phi_{V(P)}^C}\right]\, .
\end{aligned}
\end{align}
The scale parameter $\mu$, which depends on 
the masses of the initial and final states, is 
defined as 
\begin{eqnarray}\label{TCmu}
\mu=\sqrt{\Lambda_\mathrm{QCD} m_D(1-r_{P(V)}^2)}\, ,
\end{eqnarray}
for the emission diagrams ($T$ and $C$), and as
\begin{eqnarray}\label{EAmu}
\mu=\sqrt{\Lambda_\mathrm{QCD} m_D(1-r_P^2)(1-r_V^2)}\, ,
\end{eqnarray}
for the annihilation diagrams ($E$ and $A$).
Here, $ r_{P(V)}=m_{P(V)}/m_D$ denotes the mass ratio 
of the emitted or annihilating pseudoscalar (vector) meson 
from the weak vertex to the $D$ meson. The scale evolution 
of the Wilson coefficients $C_{1,2}(\mu)$ for charm quark 
decays can be found in Ref.~\cite{Li:2012cfa}. In Eq.(\ref{2amp}), 
the subscripts $PV$ and $V P$ for the amplitudes $T$ and $C$ 
indicate the order of the recoiling and emitted mesons in emission 
diagrams, where the first particle representing the recoiling meson 
and the second the emitted meson. No such distinction between 
$PV$ or $VP$ configurations applies to the annihilation diagram 
amplitudes. The superscript $V(P)$ of the effective Wilson 
coefficient $a_2$ in Eq.(\ref{a1a2}) specifies that the emitted 
meson is a vector (pseudoscalar), and the same notation 
applies to the parameters $\chi$ and $\phi$ within $a_2$.

We now illustrate the quantities appearing in the amplitude 
expressions above. The relevant CKM matrix elements 
involve the light-type quarks $q^{(')} =d,s$. The hadronic 
matrix elements are expressed in terms of the decay 
constants $f_{P}$ and $f_{V}$ of the pseudoscalar 
meson and vector meson, respectively, and the vector 
form factors  $F_{1}^{DP}$ and $A_{0}^{DV}$ for the 
$D_{(s)} \to P$ and $D_{(s)} \to V$ transitions. The 
remaining quantities, $m_V$, $\varepsilon^{*}_{V}$, 
and $p_D$, correspond to the vector meson mass, 
its polarization vector, and the $D$-meson momentum, 
respectively. The symbols 
$\chi^T,\, \chi_{V(P)}^C , \, \chi_{q(s)}^E$ , and $ \chi_{q(s)}^A $
represent the magnitudes of nonfactorizable contributions in 
the topological diagrams $T$, $C$, $E$ and $A$, respectively,
with associated strong phases denoted by 
$ \phi_{V(P)}^C , \, \phi_{q(s)}^E $, and $\phi_{q(s)}^A $.
The nonfactorizable contributions contain vertex corrections, 
hard spectator interactions involving the spectator quark of 
the $D$ meson, and FSI effects from inelastic rescattering, 
resonance effects, etc. These unknown nonfactorizable 
parameters are to be determined through a global fit to 
experimental data. To minimize free parameters, we retain 
only the relative phase of the $T$ diagram with respect to 
other amplitudes, and thus do not introduce its absolute 
strong phase. For color-suppressed $C$ diagram, 
nonfactorization contributions, such as final-state 
interaction and resonance effects, are dominated 
and differ between $C_{PV}$ and $C_{VP}$ amplitudes. 
We therefore introduce two distinct sets of parameters, 
$ \chi_{V}^C,  \phi_{V}^C$ and $ \chi_{P}^C,  \phi_{P}^C$, 
corresponding to the case where the emitted meson is a 
vector ($V$) or a pseudoscalar ($P$) particle, respectively.
For the annihilation-type $E$ and $A$ amplitudes, in 
addition to accounting for $SU(3)$ breaking effects from 
the decay constants of initial and final-state particles,
we use subscripts $q$ and $s$ in the parameters
$ \chi_{q(s)}^E, \, \chi_{q(s)}^A ,\phi_{q(s)}^E, \, $ 
and $\phi_{q(s)}^A$ to distinguish whether the strongly 
produced quark pair consists of light-quarks ($u \bar u$ 
or $d \bar d$) or strange-quarks ($s \bar s$).
As explained in Ref.~\cite{Li:2013xsa}, this treatment of 
$SU(3)$ breaking effect is more essential for accurately 
modeling the dynamics of $E$ and $A$ annihilation topologies 
in $D$ decays than accounting for isospin symmetry breaking 
($\chi_{P(V)}$ adopted in~\cite{Cheng:2010ry}). Another strong 
phase not represented in the expressions of the two annihilation 
amplitudes is the Glauber phase $ S_\pi$, which arises from 
soft gluon interactions in the two-body heavy meson decays.
It has been proposed that the Glauber eﬀect plays a significant 
role in processes with a pion in the final state~\cite{Li:2012cfa, Li:2013xsa}. 
Following Ref.~\cite{Li:2013xsa}, we incorporate this Glauber 
phase  only in the $E$ and $A$ amplitudes, but not in the 
emission diagrams. The relative phase sufficiently accounts 
for the interference effects between the annihilation and 
emission processes. In total, our parametrization of 
$D \to PV$ decays at tree level comprises 15 free 
parameters: the soft scale $\Lambda_\mathrm{QCD}$; 
 the magnitudes of the nonfactorizable amplitudes 
 $\chi^T,\, \chi_{V(P)}^C , \, \chi_{q(s)}^E$ , and 
 $ \chi_{q(s)}^A $ along with their associated strong phase 
 $ \phi_{V(P)}^C , \, \phi_{q(s)}^E $, and $\phi_{q(s)}^A $;
 and Glauber phase $ S_\pi$.

As penguin type amplitudes from Fig.~\ref{penguin} are
highly suppressed, they can be safely neglected in the 
analysis of branching ratios. However, these amplitudes 
must be retained in calculations of $\mathit{CP}$ asymmetries,
where interference between tree and penguin amplitudes 
plays a critical role. The formulas of penguin amplitudes 
are expected to be analogous in form to the tree amplitudes 
in Eq.(\ref{2amp}), but with distinct nonfactorizable parameters,
 i.e.  $\chi^{PT},\, \chi_{V(P)}^{PC} , \, \chi_{q(s)}^{PE}$ , and 
 $ \chi_{q(s)}^{PA}$ and their associated strong phases as 
 implemented in Ref.~\cite{Li:2013xsa}. In principle, these 
 parameters should be determined from experimental data, 
particularly from measurements of $\mathit{CP}$ asymmetries.
However, neither a  sufficient amount of high precision branching 
ratio data nor any measured $\mathit{CP}$ asymmetries in
two-body $D$ decays are currently available. We therefore 
adopt the approximation of using the same nonfactorizable 
parameters for the penguin amplitudes as those introduced 
for tree topologies. The treatment is well justified for the 
following reasons.

For $PT$ amplitude, the factorizable contribution dominates.
The matrix elements of the penguin operators 
$O_{3}=(\bar u_\alpha \, c_\alpha)_{V-A}\, (\bar d_\beta \, d_\beta)_{V-A}$
$O_{4}=(\bar u_\alpha \, c_\beta)_{V-A}\, (\bar d_\beta \, d_\alpha)_{V-A}$
are identical to those of the tree operators 
$O_{1}=(\bar u_\alpha \, d_\alpha)_{V-A}\, (\bar d_\beta \, c_\beta)_{V-A}$
$O_{2}=(\bar u_\alpha \, d_\beta)_{V-A}\, (\bar d_\beta \, c_\alpha)_{V-A}$
via the Fiertz transiformation.
It is therefore reasonable to adopt the same parameter $\chi^T$ 
to account for nonfactorizable effects in the $PT$ amplitudes.
The additional penguin operators 
$O_{5}=(\bar u_\alpha \, c_\alpha)_{V-A}\, (\bar d_\beta \, d_\beta)_{V+A}$ and 
$O_{6}=(\bar u_\alpha \, c_\beta)_{V-A}\, (\bar d_\beta \, d_\alpha)_{V+A}$
contribute exclusively to the penguin amplitude $PT_{VP}$, 
where the emitted particle is a pseudoscalar meson.
As an approximation, we use the same parameter $\chi^T$ to 
characterize their nonfactorizable contributions. Their explicit 
forms are given by  
\begin{align}\label{2amppt}
\begin{aligned}
PT_{PV}&=-\sqrt{2}\, G_F\, V^*_{cb}V_{ub}\, a_4(\mu)\, f_V \, m_V\, 
F_1^{DP}(m_V^2)\, (\varepsilon^{*}_{V} \cdot p_D),\\
PT_{VP}&=-\sqrt{2}\, G_F\, V^*_{cb}V_{ub}\, [a_4(\mu)-r_\chi a_6(\mu)]\, 
f_P\, m_V\, A_0^{DV}(m_P^2)\, (\varepsilon^{*}_{V} \cdot p_D)
\end{aligned}
\end{align}
with the chiral factor $r_\chi=2m^P_0/m_c$ and the effective 
Wilson coefficients
\begin{align}
\begin{aligned}
 a_4(\mu)&=C_4(\mu)+C_3(\mu)\left[{1\over N_c} + \chi^T\right],\\
a_6(\mu)&=C_6(\mu)+{C_5(\mu)} \left[{1\over N_c}+\chi^{T} \right].
\end{aligned}
\end{align}
Since the penguin annihilation diagram $PA$ share the 
same topological structure as $PT$ and differ only in gluon 
interactions, it can be regarded as a power correction to $PT$ 
amplitude. Its contribution is therefore absorbed into the 
nonfactorizable parameter of $PT$, specifically, through 
the parameter $\chi^T$, as done in Ref.~\cite{Zhou:2016jkv}.

The nonfactorizable contribution of color-suppressed $PC$ 
amplitude form $O_{4}$ and $O_{6}$ are found to be identical 
in the PQCD analyses~\cite{Zou:2013ez}. We thus use the
same parameters $\chi_{V(P)}^{C}$ and $\phi_{V(P)}^{C}$ 
for $PC$ amplitude, as given below
\begin{align}\label{2amppc}
\begin{aligned}
 PC_{PV}&=-\sqrt{2}\, G_F\, V^*_{cb}V_{ub}\, \left[a_3^{V}(\mu)+a_5^{V}(\mu)\right]\, 
f_V\, m_V\, F_1^{DP}(m_V^2)\, (\varepsilon^{*}_{V} \cdot p_D),\\
PC_{VP}&=-\sqrt{2}\, G_F\, V^*_{cb}V_{ub}\, \left[a_3^P(\mu)-a_5^P(\mu)\right]\, 
f_P\, m_V\, A_0^{DV}(m_P^2)\, (\varepsilon^{*}_{V} \cdot p_D),
\end{aligned}
\end{align}
with the effective Wilson coefficients
\begin{align}
\begin{aligned}
a_3^{V(P)}(\mu)&=C_3(\mu)+C_4(\mu) \left[{1\over N_c}+
\chi_{V(P)}^{C} \mathrm {e}^{i\, \phi_{V(P)}^{C} }\right] \, ,\\
a_5^{V(P)}(\mu)&=C_5(\mu)+C_6(\mu) \left[{1\over N_c}-
\chi_{V(P)}^{C} \mathrm {e}^{i\, \phi_{V(P)}^{C} }\right]\, .
\end{aligned}
\end{align}

For the amplitude $PE$, the nonfactorizable contributions 
from operators $O_{4,6}$ have also been analyzed in 
Ref.~\cite{Zou:2013ez} and can be expressed using 
the same parameterization as
\begin{align}\label{2amppe}
 PE_{PV, VP}&=-\sqrt{2}\, G_F\, V^*_{cb}V_{ub}\, 
[C_4(\mu)\,  -\, C_6(\mu) ]\, 
 \chi^{E}_{q,s} \, \mathrm{e}^{i\phi^{E}_{q,s}}\,
f_D\, m_V \, {f_P\over f_\pi}{f_V\over f_\rho}\,  
(\varepsilon^{*}_{V} \cdot p_D)\, .
\end{align}

In the subsequent analysis of {\it CP} violation, we also 
incorporate contributions from chromomagnetic penguin 
operators and quark-loops corrections, which influence 
penguin-induced {\it CP} asymmetries. Following the 
treatment in Ref.~\cite{Beneke:1999br,Beneke:2000ry,Beneke:2003zv}, 
these effects are absorbed into an effective redefinition of 
the Wilson coefficients $C_{3,4,5,6}$ and explicit expressions 
can also be found in Ref.~\cite{Li:2013xsa}.
\section{Numerical Results and discussions}\label{sec:3}
\subsection{Input parameters}
The input parameters in the amplitudes of Eq.(\ref{2amp}) can be
classified as electroweak coefficients, short-distance physics coefficients 
and nonperturbative QCD parameters. 
The electroweak parameters, namely the CKM matrix elements, are taken from
the PDG~\cite{ParticleDataGroup:2024cfk}. The Wilson coefficients,
encoding the short-distance physics in $D$ decays, are 
adopted from the Appendix of Ref.~\cite{Li:2012cfa}.

The nonperturbative QCD input include decay constants and transition form factors.
The decay constants for pseudoscalar mesons and vector mesons 
 (in units of MeV) are summarized in Table~\ref{tab:decay constants}. 
 Those for the $\pi, K$ and $D_{(s)}$ are taken from the 
 PDG~\cite{ParticleDataGroup:2024cfk}, while for vector mesons 
 not yet measured experimentally, we employ the same theoretical values 
 used in quasi-two-body decays of $B$-meson, 
 assigning 5\% uncertainty~\cite{Zhou:2024qmm}.
\begin{table}[tbhp]
\caption{ Decay constants of pseudoscalar and vector mesons. }
\vspace{3mm}
\label{tab:decay constants}
\centering
\begin{tabular}{ccccccccccc|}
\hline
$f_{\pi}$ & $f_{K}$  &  $f_{D}$ & $f_{D_s}$ &$f_{\rho}$ &$f_{K^*}$ & $f_{\omega}$&  $f_{\phi}$ &
\\ \hline
$130.2 \pm 1.7$ & $155.6 \pm 0.4$ & $211.9 \pm 1.1$ & $258 \pm 12.5$ & $213 \pm 11$ & $220 \pm 11$ & $192\pm 10 $ & $225\pm11$ &
\\
\hline
\end{tabular}
\end{table}

The transition form factors for $D$ decays, such as $D \to \pi$ and $D \to K$,
have been measured by experiments including CLEO-c~\cite{CLEO:2009svp},
Belle~\cite{Belle:2006idb}, and BESIII~\cite{BESIII:2018xre, BESIII:2024zft}.
For transitions lacking experimental measurements, 
theoretical predictions are available from a variety of approaches,
such as Lattice QCD~\cite{Bernard:2009ke}, 
QCD sum rules~\cite{Du:2003ja,Tian:2024lrn},
quark models~\cite{Melikhov:2000yu}, 
the covariant light-front quark model~\cite{Chen:2009qk,Verma:2011yw,Wang:2008ci},
Improved light-cone harmonic oscillator model~\cite{Hu:2024tmc},
heavy meson and chiral symmetries ($\mathrm{HM_\chi T}$)~\cite{Fajfer:2005ug},
Hard-Wall AdS/QCD model~\cite{Momeni:2022gqb},
and 4-flavor holographic QCD~\cite{Ahmed:2023pod}.
In light of the discrepancies in form factor predictions from different 
approaches, the values adopted at zero recoil ($Q^2=0$) are 
chosen to be consistent with the current experimental and theoretical landscape.
A 10 \% uncertainty is consequently kept to
account for the theoretical variations. 
The adopted values of the form factors at $Q^2=0$ and the associated 
$\alpha_i$ parameters from Ref.~\cite{Fu-Sheng:2011fji}, 
are provided in Table~\ref{tab:ff}.
Their $Q^2$ dependence is uniformly parameterized by the dipole form used 
in Ref.~\cite{Fu-Sheng:2011fji,Li:2012cfa, Li:2013xsa}
 \begin{equation}\label{eq:ffdipole}
F_{i}(Q^{2})
={F_{i}(0)\over 1-\alpha_{1}{Q^{2}\over m_{\rm pole}^{2}}+\alpha_{2}{Q^{4}\over m_{\rm pole}^{4}}},
\end{equation}
where $F_{i}$ stands for $F_{1}$ or $A_{0}$. The pole mass $m_{\rm pole}$ takes the value  
$m_{D^*}$ for $F_{1}^{D \pi, D \eta^{(')}, D_s K}$, 
$m_{D_s^{*}}$ for $F_{1}^{DK,D_s \eta^{(')}}$, 
$m_D$ for $A_{0}^{D \rho, D \omega, D_s K^*}$,
and $m_{D_s}$ for $A_{0}^{D K^*, D_s \phi}$. 
\begin{table} [hbt]
\caption{Form factors and dipole model parameters. }\label{tab:ff}
\vspace{3mm}
\centering
\begin{tabular}{|c|c|c|c|c|c|c|c|c|c|c|}
\hline&
$~~F_{1}^{D\to\pi}~~$&
$~~F_{1}^{D\to K}~~$&
$~~F_{1}^{D_s\to K}~~$&
$~~F_{1}^{D\to \eta_q}~~$&
$~~F_{1}^{D\to \eta_s}~~$\\
\hline
$~F_i(0)~$&
0.60&
0.74&
0.66 &
0.76&
0.79\\
$\alpha_1$&
1.24&
1.33&
1.20&
1.03&
1.23\\
$\alpha_2$&
0.24&
0.33&
0.20&
0.29&
0.23\\
\hline&
$~~A_{0}^{D\to \rho}~~$&
$~~A_{0}^{D\to K^*}~~$ &
$~~A_{0}^{D_s\to K^*}~~$ &
$~~A_{0}^{D\to \omega}~~$&
$~~A_{0}^{D_s \to \phi}~~$ \\
\hline
$~F_i(0)~$&
0.84&
0.73&
0.68&
0.68&
 0.70  \\
$\alpha_1$&
1.36&
1.17&
1.20&
1.36&
 1.10 \\
$\alpha_2$&
0.36&
0.17&
0.20&
0.36&
0.10 \\
\hline
\end{tabular}
\end{table}

\subsection{The updated nonfactorizable parameters}
The nonfactorizable parameters of the topological tree diagram 
amplitudes of Eq.(\ref{2amp}) consist of 15 free quantities:
$\chi^T,\, \chi_{V(P)}^C , \, \chi_{q(s)}^E, \, \text{and}  \, \chi_{q(s)}^A $, 
along with their associated phases $\phi_{V(P)}^C , \, \phi_{q(s)}^E, \, \phi_{q(s)}^A \, $, 
and $S_\pi$ (where the factor $\mathrm{e}^{i\, S_\pi}$ multiplies the $E$ and $A$ amplitudes),  
as well as the soft scale $\Lambda_{\mathrm{QCD}}$ [see Eqs.(\ref{TCmu}) (\ref{EAmu})]. 
These unknown parameters are determined by a global fit to 41 experimental 
measurements of $D \to PV$ branching ratios with significance above $3 \sigma$
from PDG~\cite{ParticleDataGroup:2024cfk}.
This dataset has higher precision and includes more data points than the 33 used in~\cite{Li:2013xsa}.
The best-fit values and their corresponding uncertainties are as follows
\begin{align}\label{parameter}
\begin{aligned}
\chi^{T}&=-0.29 \pm 0.01\, ,~~~ \\
\chi_P^{C}&=-0.47 \pm 0.02\, ,~~~~\phi_P^{C}=0.37 \pm 0.12,\\
\chi_V^{C}&=-0.41 \pm 0.01\, ,~~~~\phi_V^{C}=-0.52\pm 0.02, \\
\chi_q^{E}&=0.13\pm0.01\, ,~~~~~~\phi_q^{E}=2.68\pm 0.10, \\
\chi_s^{E}&=0.24\pm0.01\, ,~~~~~~\phi_s^{E}=3.52\pm 0.07, \\
\chi_q^{A}&=0.10\pm0.003\, ,~~~~\phi_q^{A}=-2.08\pm 0.16, \\
\chi_s^{A}&=0.18\pm0.01\, ,~~~~~~\phi_s^{A}=2.62\pm 0.10, \\
S_\pi&=-1.80\pm 0.17\, ,\, \Lambda_\mathrm{QCD}=(0.24\pm0.01) \mathrm{GeV}\, ,
\end{aligned}
\end{align}
with $\chi^2/\mathrm{d.o.f.}=6.2$.
These nonfactorizable parameters are determined with high
precision, with the exception of the strong phase $\phi_P^{C}$.
The precision for this phase is limited due to insufficient experimental data
from decay modes dominated by the $C_{VP}$ amplitude.
The nonfactorizable contribution of the $T$ topology, $\chi^{T}$, is non-negligible.
 The term $C_1\,  \chi^{T}$ is comparable in magnitude to its factorizable counterpart,  
 $ C_1/N_C$, from $O_1$ operator.
The $C$ amplitude parameters $\chi_P^{C}$ and  $\chi_V^{C}$ show 
slight deviations from the values obtained early~\cite{Li:2013xsa} 
(note that the labeling of $\chi_P^{C}$ and $\chi_V^{C}$ is reversed 
relative to Ref.~\cite{Li:2013xsa}).
The more significant discrepancies, however, arises from the 
the $E$ and $A$ topologies, which are dominated by 
nonfactorizable dynamics. Specifically, the fitted values of $\chi_{q,s}^{E,A}$
together with the strong phases $\phi_{q,s}^{E,A}$ and the 
Glauber phase $S_\pi$, collectively account for these observed differences.
The reason lie in the fact  that the dominant $T$ and $C$ amplitudes 
were already well-determined with the prior data, 
whereas the subleading $E$ and $A$ topological amplitudes 
are smaller and thus require a larger set of high-precision
measurements to be accurately constrained.
The $\Lambda_\mathrm{QCD}$ obtained in this work is physically reasonable.
For instance, in the $D^+ \to \bar K^0 \, \rho^+$ decay mode,
the corresponding scale is $\mu=0.61\, \mathrm{GeV}$ 
(with $\Lambda_\mathrm{QCD}=0.24\, \mathrm{GeV}$),
yielding a value approximately $m_c/2$.
At this scale, the Wilson coefficients for
the $T$ and $C$ amplitudes 
are $a_1(\mu)=1.45$ and $a_2^P(\mu)=-1.09 \, \mathrm{e}^{i\, 0.26}$,
indicating that while the $T$ topology dominates $D$ decays,
the $C$ topology also features substantial nonfactorizable effects.
\subsection{Branching Ratios}
Our numerical results for the branching ratios of $D_{(s)}$ decays 
are collected in Tables \ref{BrCF}-\ref{BrDCS}, 
corresponding to the CF ($V_{cs}^*\, V_{ud}$), 
SCS ($V_{cd(s)}^*\, V_{ud(s)}$), and 
DCS ($V_{cd}^*\, V_{us}$) modes , respectively. 
In our results, denoted as `` Br(FAT)", the first uncertainty
is statistical, originating from the nonperturbative parameters 
in Eq.(\ref{parameter}) via the $\chi^2$ fit to the experimental data.
The second and third uncertainties arise from the form factors 
and the decay constants, respectively. 
As shown in our results, the uncertainty is dominated by 
the contributions from the form factors.
The topological diagram amplitudes $T_{PV,VP}, C_{PV,VP}, E, A$
are provided in the second columns of these tables
to facilitate the analysis of branching ratio hierarchies therein.
We list all available experimental data, including those
with significance below $3 \sigma$,
in the third column for comparing with our results.
Finally, the last two columns present the latest  results 
updated in the topological diagram approach for comparison, 
specifically the  (F4) and $\mathrm{(F1^\prime)}$
corresponding to two distinct solutions derived from different 
fitted topological amplitude parameters~\cite{Cheng:2024hdo}.
Our results are consistent with the measured CF and SCS modes.
Regarding DCS decays, most of them have not yet been 
experimentally measured, so our predictions can be tested 
with future data. 
Compared with the results from the topological diagram approach,
most of our findings are in good agreement with them.
The updated results in the two different 
theoretical approaches, especially for DCS decays,  
can serve as a valuable cross-check.
\begin{table}[!htbh]
  \centering
  \caption{Branching ratios for the Cabibbo-favored $D\to PV$ decays
  in units of percentage. Our results (FAT) are compared with
  the experimental data (EXP)~\cite{ParticleDataGroup:2024cfk} and 
  the updated results  (F4) and  $\mathrm{(F1^\prime)}$ from the topological diagram approach.
The second column lists the corresponding topological diagram contributions,
denoted by $T_{PV,VP}, C_{PV,VP}, E$ and $A$.} 
 \label{BrCF}
 \begin{adjustbox}{width=\textwidth}
 \begin{tabular}[t]{lccccc}\hline\hline
  Modes & Amplitudes & Br(EXP)   &  Br(FAT) &  Br(F4)&  $\mathrm{Br(F1^\prime)}$\\\hline
  $D^0\to \pi^+K^{*-}$ & $T_{VP},E$             & $5.34\pm0.41$   &$6.05\pm0.46\pm0.97\pm0.17$  & $5.45\pm0.34$ & $5.37\pm0.33$\\
  $D^0\to \pi^0\overline K^{*0}$ & $C_{PV}, E$& $3.74\pm0.27$ &$3.33\pm0.30\pm0.58\pm0.34$  & $3.61\pm0.18$ & $3.70\pm0.18$\\
  $D^0\to\ \overline{K}^0\rho^0$ & $C_{VP}, E$  & $1.26\pm0.16$  &$1.08\pm0.14\pm0.27\pm0.04$ & $1.25\pm0.09$ & $1.25\pm0.09$\\
  $D^0\to\ \overline{K}^0\omega$ & $C_{VP}, E$  & $2.22\pm0.12$ &$2.06\pm0.16\pm0.31\pm0.08$ & $2.29\pm0.11$ & $2.29\pm0.11$ \\
  $D^0\to\ \overline{K}^0\phi$ & $E$            & $0.825\pm0.061$   &$0.64\pm0.06\pm0.00\pm0.09$ & $0.830\pm0.034$ & $0.828\pm0.034$\\
  $D^0\to\ K^-\rho^+$ & $T_{PV}, E$            & $11.2\pm0.7$  &$10.41\pm0.48\pm2.46\pm1.06$  & $11.4\pm0.8$ & $11.3\pm0.7$\\

  $D^0\to\ \eta\overline K^{*0}$ & $C_{PV}, E$  & $1.41\pm0.12$   &$0.92\pm0.11\pm0.18\pm0.09$  & $1.35\pm0.06$ & $1.41\pm0.07$\\

  $D^0\to\ \eta'\overline K^{*0}$ & $C_{PV}, E$ & $\leq 0.1$ &$0.013\pm0.002\pm0.001\pm0.001$  &$0.0055\pm0.0004$ &$0.0043\pm0.0003$  \\

  $D^+\to\ \pi^+\overline K^{*0}$ & $T_{VP}$,$C_{PV}$ & $1.57\pm0.13$  &$1.59\pm0.13\pm0.51\pm0.24$  & $1.58\pm0.13$ & $1.58\pm0.13$ \\

  $D^+\to\ \overline{K}^0\rho^+$ & $T_{PV}$,$C_{VP}$  &$12.28\pm1.2$  &$10.48\pm1.11\pm4.20\pm1.92$  & & \\

  $D_s^+\to\ \pi^+\rho^0$ & $A$        & $0.020\pm0.012$  & 0    & $0.011\pm0.003$    & $0.011\pm0.002$                     \\

  $D_s^+\to  \pi^+\omega$ & $A$       & $0.238\pm0.015$   &$0.24\pm0.02\pm0.00\pm0.03$ & $0.24\pm0.01$ & $0.24\pm0.01$ \\

  $D_s^+\to\ \pi^+\phi$ & $T_{VP}$                & $4.5\pm0.12$  &$4.45\pm0.08\pm0.89\pm0.01$  & $4.49\pm0.11$ & $4.50\pm0.11$ \\

  $D_s^+\to\ \pi^0\rho^+$ & $A$        &  & 0 & $0.012\pm0.003$ & $0.011\pm0.002$\\

  $D_s^+\to\ K^+\overline K^{*0}$ & $C_{PV}, A$ & $3.79\pm0.09$  &$3.70\pm0.25\pm1.00\pm0.39$   & $3.80\pm0.10$  & $3.79\pm0.09$\\

  $D_s^+\to\ \overline{K}^0K^{*+}$ & $C_{VP}, A$ & $1.54\pm0.14$ &$1.72\pm0.24\pm0.42\pm0.05$  & & \\

  $D_s^+\to\ \eta\rho^+$ & $T_{PV}, A$  & $8.9\pm0.8$  &$9.10\pm0.39\pm1.91\pm0.91$ & $9.25\pm0.35$  & $8.75\pm0.31$ \\

  $D_s^+\to\ \eta'\rho^+$ & $T_{PV}, A$ &$5.8\pm1.5$ &$4.04\pm0.11\pm0.77\pm0.40$ & $3.24\pm0.11$ & $3.60\pm0.11$
   \\
  \hline\hline
  \end{tabular}
\end{adjustbox}
\end{table}

\begin{table}[!htbh]
  \centering
  \caption{Same as Table~\ref{BrCF} for the singly Cabibbo-suppressed
  $D\to PV$ decays in units of $10^{-3}$.}  \label{BrSCS}
\scalebox{0.88}{
 \begin{tabular}[t]{lccccc}\hline\hline
  Modes & Amplitudes & Br(EXP)   &  Br(FAT) &  Br(F4)&  $\mathrm{Br(F1^\prime)}$ \\ \hline
  $D^0\to \pi^+\rho^-$ & $T_{VP}, E$               & $5.15\pm0.25$  &$4.93\pm0.34\pm0.82\pm0.12$   & $5.42\pm0.12$  & $5.23\pm0.18$  \\

  $D^0\to \pi^0\rho^0$ &  $C_{PV}$,$C_{VP}, E$& $3.86\pm0.23$ &$3.04\pm0.28\pm0.34\pm0.21$  & $2.86\pm0.06$  & $3.38\pm0.10$ \\

  $D^0\to \pi^0\omega$ & $C_{PV}$,$C_{VP}, E$   &$0.117\pm0.035$  &$0.23\pm0.08\pm0.08\pm0.05$ &$0.157\pm0.015$ &$0.153\pm0.021$\\

  $D^0\to \pi^0\phi$ & $C_{PV}$                   &  $1.17\pm0.04$    &$1.16\pm0.05\pm0.23\pm0.12$ &$0.93\pm0.02$  &$0.99\pm0.02$\\

  $D^0\to \pi^-\rho^+$ & $T_{PV}, E$            & $10.1\pm0.4$   &$9.65\pm0.51\pm1.70\pm0.97$ & $10.6\pm0.5$  & $10.2\pm0.6$ \\

  $D^0\to K^+K^{*-}$ & $T_{VP}, E$            &$1.65\pm0.11$    & $1.30\pm0.10\pm0.34\pm0.06$   &$1.65\pm0.04$ &$1.55\pm0.04$\\

  $D^0\to K^0\overline{K}^{*0}$ & $E$       &$0.246\pm0.048$ &$0.35\pm0.07\pm0.00\pm0.05$  &$0.246\pm0.011$ &$0.246\pm0.021$\\

  $D^0\to \overline{K}^0K^{*0}$ & $E$      &$0.336\pm0.063$  &$0.35\pm0.07\pm0.00\pm0.05$  &$0.336\pm0.021$ &$0.336\pm0.015$ \\

  $D^0\to K^-K^{*+}$ & $T_{PV}, E$          &$4.56\pm0.21$    & $5.01\pm0.21\pm1.17\pm0.51$  &  $4.57\pm0.22$  &  $4.56\pm0.15$\\

  $D^0\to \eta\rho^0$ & $C_{PV}$,$C_{VP}, E$&  &$0.49\pm0.12\pm0.10\pm0.00$  &$0.26\pm0.02$ &$0.25\pm0.02$ \\

  $D^0\to \eta\omega$ & $C_{PV}$,$C_{VP}, E$&$1.98\pm0.18$ &$2.78\pm0.16\pm0.35\pm0.13$  &$1.71\pm0.05$ &$1.99\pm0.06$\\

  $D^0\to \eta\phi$ & $C_{PV}, E$          &$0.18\pm0.05$  & $0.25\pm0.05\pm0.01\pm0.03$ &$0.175\pm0.007$  &$0.186\pm0.004$   \\

  $D^0\to \eta'\rho^0$ & $C_{PV}$,$C_{VP}, E$ & & $0.24\pm0.01\pm0.03\pm0.01$ &$0.059\pm0.002$  &$0.059\pm0.002$  \\

  $D^0\to \eta'\omega$ & $C_{PV}$,$C_{VP}, E$ & & $0.02\pm0.00\pm0.00\pm0.00$  &$0.017\pm0.001$ &$0.009\pm0.001$       \\

  $D^+\to \pi^+\rho^0$ & $T_{VP}$,$C_{PV}, A$ &$0.83\pm0.14$ & $0.51\pm0.05\pm0.13\pm0.01$  &$0.55\pm0.06$  &$0.57\pm0.05$  \\
  $D^+\to \pi^+\omega$ & $T_{VP}$,$C_{PV}, A$ & $0.28\pm0.06$  & $0.28\pm0.10\pm0.19\pm0.05$  & $0.31\pm0.05$  & $0.88\pm0.07$\\

  $D^+\to \pi^+\phi$ & $C_{PV}$                       &$5.7\pm0.14$ &$5.91\pm0.22\pm1.18\pm0.59$  & $4.74\pm0.10$ & $5.03\pm0.10$\\

  $D^+\to \pi^0\rho^+$ & $T_{PV}$,$C_{VP}, A$ &  &$3.20\pm0.28\pm1.18\pm0.55$ &$5.20\pm0.33$ &$5.25\pm0.38$ \\

  $D^+\to K^+\overline{K}^{*0}$ & $T_{VP}, A$    & $3.71\pm0.18$  &$3.91\pm0.20\pm0.94\pm0.12$ & $5.78\pm0.15$ & $5.26\pm0.14$\\

  $D^+\to \overline{K}^0K^{*+}$ & $T_{PV}, A$  & $17.3\pm1.8$ &$13.98\pm0.42\pm3.12\pm1.41$   & $15.8\pm0.5$  & $15.6\pm0.5$\\

  $D^+\to \eta\rho^+$ & $T_{PV}$,$C_{VP}, A$ & & $0.18\pm0.16\pm0.13\pm0.04$  & $0.38\pm0.18$ & $0.36\pm0.19$ \\

  $D^+\to \eta'\rho^+$ & $T_{PV}$,$C_{VP}, A$& & $1.71\pm0.05\pm0.24\pm0.10$  & $0.97\pm0.03$ & $1.12\pm0.03$ \\

  $D_s^+\to \pi^+K^{*0}$ & $T_{VP}, A$    &$2.55\pm0.35$ &$2.28\pm0.16\pm0.51\pm0.03$ &$2.06\pm0.06$ &$1.58\pm0.05$\\

  $D_s^+\to \pi^0K^{*+}$ & $C_{VP}, A$        &$0.75\pm0.25$ &$0.51\pm0.07\pm0.11\pm0.01$ &$0.71\pm0.03$ &$0.67\pm0.03$ \\

  $D_s^+\to K^+\rho^0$ & $C_{PV}, A$          & $2.17\pm0.25$ &$1.84\pm0.11\pm0.37\pm0.18$    & $1.01\pm0.03$  & $1.11\pm0.03$   \\
  $D_s^+\to K^+\omega$ & $C_{PV}, A$         & $0.99\pm0.15$ & $1.51\pm0.09\pm0.29\pm0.15$ & $1.66\pm0.03$ & $1.17\pm0.03$\\

  $D_s^+\to K^+\phi$ & $T_{VP}$,$C_{PV}, A$      & $0.18\pm0.04$   & $0.24\pm0.05\pm0.11\pm0.03$ & $0.11\pm0.01$ & $0.29\pm0.02$\\

  $D_s^+\to K^0\rho^+$ & $T_{PV}, A$         &$5.46\pm0.95$ &$9.35\pm0.33\pm1.95\pm0.94$  &$7.54\pm0.27$    &$7.30\pm0.26$\\

  $D_s^+\to \eta K^{*+}$ & $T_{PV}$,$C_{VP}, A$ &  & $1.52\pm0.27\pm0.52\pm0.28$ &$0.37\pm0.07$ &$0.39\pm0.09$\\

  $D_s^+\to \eta'K^{*+}$ & $T_{PV}$,$C_{VP}, A$ & &$0.68\pm0.03\pm0.16\pm0.09$ &$0.40\pm0.02$ &$0.42\pm0.02$\\
  \hline\hline
  \end{tabular}
  }
\end{table}

\begin{table}[!htbh]
  \centering
  \caption{Same as Table~\ref{BrCF} for the doubly Cabibbo-suppressed
  $D\to PV$ decays in units of $10^{-4}$.}  \label{BrDCS}
   \begin{adjustbox}{width=\textwidth}
   \begin{tabular}[t]{lccccc}\hline\hline
  Modes & Amplitudes & Br(EXP)   &  Br(FAT) &  Br(F4)&  $\mathrm{Br(F1^\prime)}$ \\\hline
  $D^0\to \pi^0K^{*0}$ & $C_{PV}, E$       &  &$0.95\pm0.08\pm0.16\pm0.10$ &$0.84\pm0.04$ &$0.48\pm0.02$ \\

  $D^0\to \pi^-K^{*+}$ & $T_{PV}, E$       & $3.39^{+1.80}_{-1.02}$&$4.91\pm0.24\pm0.87\pm0.49$  &$3.54\pm0.28$ &$3.46\pm0.17$\\

  $D^0\to K^+\rho^-$ & $T_{VP}, E$           & &$1.41\pm0.09\pm0.35\pm0.05$  &$1.30\pm0.07$ &$1.32\pm0.04$\\

  $D^0\to K^0\rho^0$ & $C_{VP}, E$          & &$0.31\pm0.04\pm0.08\pm0.01$  &$0.25\pm0.02$ &$0.27\pm0.01$\\

  $D^0\to K^0\omega$ & $C_{VP}, E$         &  &$0.59\pm0.05\pm0.09\pm0.02$  &$0.66\pm0.03$ &$0.51\pm0.02$\\

  $D^0\to K^0\phi$ & $E$                   &   &$0.18\pm0.02\pm0.00\pm0.03$ &$0.22\pm0.01$ &$0.55\pm0.01$\\

  $D^0\to \eta K^{*0}$ & $C_{PV}, E$        & &$0.26\pm0.03\pm0.05\pm0.03$  &$0.34\pm0.02$ &$0.20\pm0.01$\\

  $D^0\to \eta'K^{*0}$ & $C_{PV}, E$   & &$0.0037\pm0.0005\pm0.0004\pm0.0004$ &$0.0019\pm0.0001$ &$0.0016\pm0.0001$ \\

  $D^+\to \pi^+K^{*0}$ & $C_{PV}, A$       &$3.45\pm0.6$   &$3.07\pm0.12\pm0.68\pm0.31$ &$2.52\pm0.07$ &$2.51\pm0.06$ \\

  $D^+\to \pi^0K^{*+}$ & $T_{PV}, A$       & $3.4\pm1.4$ &$4.54\pm0.14\pm0.95\pm0.45$ &$4.17\pm0.15$ &$4.12\pm0.14$ \\

  $D^+\to K^+\rho^0$ & $T_{VP}, A$        &$1.9\pm0.5$    &$2.63\pm0.10\pm0.55\pm0.03$ &$1.84\pm0.05$ &$1.54\pm0.04$ \\

  $D^+\to K^+\omega$ & $T_{VP}, A$        &  $0.57^{+0.25}_{-0.21}$ &$2.10\pm0.07\pm0.40\pm0.03$ &$2.09\pm0.05$ &$2.42\pm0.05$ \\

  $D^+\to K^+\phi$ & $A$                    &$0.09\pm0.012$   &$0.07\pm0.01\pm0.00\pm0.01$   &$0.057\pm0.002$ &$0.032\pm0.003$  \\

  $D^+\to K^0\rho^+$ & $C_{VP}, A$         &  &$2.17\pm0.18\pm0.48\pm0.04$  &$1.40\pm0.06$ &$1.38\pm0.05$\\

  $D^+\to \eta K^{*+}$ & $T_{PV}, A$   & $4.4^{+1.80}_{-1.50} $ &$1.86\pm0.08\pm0.41\pm0.19$ &$1.44\pm0.05$ &$1.29\pm0.04$\\

  $D^+\to \eta'K^{*+}$ & $T_{PV}, A$   & &$0.03\pm0.00\pm0.00\pm0.00$ &$0.016\pm0.001$ &$0.021\pm0.001$\\

  $D_s^+\to K^+K^{*0}$ & $T_{VP}$,$C_{PV}$          & $0.90\pm0.50$  &$0.22\pm0.02\pm0.03\pm0.02$ &$0.20\pm0.02$ &$0.11\pm0.02$\\
  $D_s^+\to K^0K^{*+}$ & $T_{PV}$,$C_{VP}$          &  &$1.96\pm0.14\pm0.67\pm0.31$  &$1.47\pm0.09$ &$1.44\pm0.11$\\
  \hline\hline
  \end{tabular}
  \end{adjustbox}
\end{table}

Among the CF modes, the decays $D^0 \to K^- \rho^+$ 
(with $T_{PV}$ and $E$ amplitudes) and 
$D^+ \to \bar K^0 \, \rho^+$ (with $T_{PV}$ and $C_{VP}$ amplitudes)
exhibit the largest branching ratios, confirming their dominance by 
the $T_{PV}$ topology.
These largest branching ratios originate from the large 
Wilson coefficient $a_1(\mu)$ and sizable vector meson 
decay constant in $T_{PV}$ amplitude. 
Although the ratio $|a_2^{P}/a_1|=0.75$ might suggest comparable
contributions from the $T_{PV}$ and $C_{VP}$ in $D^+ \to \bar K^0 \, \rho^+$,
the smaller pseudoscalar meson decay constant in $C_{VP}$ amplitude 
leads to a dominant $T_{PV}$ contribution and a suppressed $C_{VP}$ 
component.
In contrast, for $D^+ \to \pi^+ \, \bar K^{*0}$,  the $T_{VP}$ and $C_{PV}$ 
amplitudes add incoherently, yielding a branching ratio nearly one order 
of magnitude smaller than that of $D^+ \to \bar K^0 \, \rho^+$.
In decays involving both the $T$ and $E$ or $C$ and $E$ topologies,
a strong coherence cannot be established between the amplitude 
$T_{PV\, (VP)}$ and $E$ or between $C_{PV\, (VP)}$ and $E$.
This is supported by the decay $D^0 \to \pi^0 \, \bar K^{*0}$, where
the ratio $E/C_{PV}=0.18$ implies a strong suppression of the $E$ amplitude.
Therefore, the $E$ diagram can be neglected to a good approximation 
in the processes involving both $T$ and $E$, or $C$ and $E$. 
 The observed branching fraction relations in CF decays,
 $\mathrm{Br}(D^0 \to K^- \rho^+)\,  \approx\,  2 \,  \mathrm{Br}(D^0 \to \pi^+ K^{*-})$ 
 $\mathrm{Br}(D^0 \to \pi^0 \bar K^{*0})\,  \approx\,  3 \,  \mathrm{Br}(D^0 \to \bar K^0 \rho^0 )$,
 and $\mathrm{Br}(D_s^+ \to K^+ \bar K^{*0})\,  \approx\,  2 \,  \mathrm{Br}(D^0 \to \bar K^0 K^{*+} )$
 can be explained by the negligible $E$ amplitudes and the consequent 
 by a factor of $2-3$ for modes with $T_{PV}$ or $C_{PV}$ amplitudes
 over those with $T_{VP}$ or $C_{VP}$ modes, owing to the much larger 
 the decay constants of light vector mesons
 compared to light pseudoscalar ones.
 A similar hierarchy  is likewise evident in SCS decays,
 $\mathrm{Br}(D^0 \to \pi^- \rho^+)\,  \approx\,  2 \,  \mathrm{Br}(D^0 \to \pi^+ \rho^-)$,
 $\mathrm{Br}(D^0 \to K^- K^{*+})\,  \approx\,  3 \,  \mathrm{Br}(D^0 \to K^+ K^{*-})$,
 $\mathrm{Br}(D^+ \to \bar K^0 K^{*+})\,  \approx\,  3 \,  \mathrm{Br}(D^+ \to K^+ \bar K^{*0})$,
 and $\mathrm{Br}(D_s^+ \to K^+ \rho^0)\,  \approx\,  3 \,  \mathrm{Br}(D_s^+ \to \pi^0  K^{*+})$,
 and DCS decays,
 $\mathrm{Br}(D^0 \to \pi^- K^{*+})\,  \approx\,  3 \,  \mathrm{Br}(D^0 \to K^+ \rho^-)$,
 $\mathrm{Br}(D^+ \to \pi^0 K^{*+})\,  \approx\,  2 \,  \mathrm{Br}(D^0 \to K^+ \rho^0)$.
 
  In the FAT approach, the branching ratios of 
  $D_s^+ \to \pi^+\, \rho^0$ and $D_s^+ \to \pi^0\, \rho^+$,
which receive contributions solely from $A$ topological diagram,
are predicted to be zero. This arises because we do not introduce 
two distinct parameters $\chi_P^A$ and $\chi_V^A$ to account for
nonfactorizable contributions where the spectator antiquark 
enters a pseudoscalar ($P$) or a vector ($V$) mesons, respectively. 
Consequently, the contributions from the $A$ diagram involving 
$ u \, \bar u$ forming $\pi^0 (\rho^0)$ meson 
 and $ d \, \bar d$ forming $ \rho^0 (\pi^0)$ cancel exactly.
 As demonstrated in~\cite{Li:2013xsa}, SU(3) symmetry breaking play a more crucial 
role than the isospin symmetry breaking (which would be parameterized by 
$\chi_P^A$ and $\chi_V^A$)  in $D$ meson decays.
 We thus employ two parameters, $\chi^A_q$ and $\chi^A_s$, to 
capture SU(3) symmetry breaking effects, rather than introducing
additional parameters for subleading isospin symmetry breaking.


\subsection{ Direct $\mathit{CP}$ asymmetries}
With the fitted parameters in Eq.(\ref{parameter}), we calculate 
the direct $\mathit{CP}$ asymmetries for the $D \to PV$, as shown 
in Table~\ref{directCP}. The second column in this table lists 
the penguin topological diagram amplitudes ($PT, PC$, and 
$PE$) in the absence of the tree amplitudes. 
The direct $\mathit{CP}$ asymmetries in the third column result from
the interference between tree and penguin amplitudes.
Incorporating contributions from chromomagnetic penguins (cm) 
and quark loops (ql), which have been absorbed into the 
penguin Wilson coefficients as specified in 
Eqs.(\ref{2amppt}-\ref{2amppe}), yields the final results 
 presented in the fourth column. 
The uncertainties in the $\mathit{CP}$ asymmetries originate from
the same sources as those in branching ratios,
but are now dominated by nonfactorizable parameters 
due to the significant cancellation of hadronic uncertainties from 
form factors and decay constants.
The updated strong phases in Eq.(\ref{parameter}) lead to
$\mathit{CP}$ asymmetries that are markedly different from 
previous results~\cite{Li:2013xsa}.
In contrast to a recent topological diagram 
approach~\cite{Cheng:2021yrn}, our results within the FAT approach 
show significant differences.
This discrepancy arises because that study incorporates 
penguin contributions based on QCD factorization approach, 
and only the nonfactorizable effects from $PE$ topology are included.
\begin{table}[tbhp]
\caption{Direct $\mathit{CP}$ asymmetries for the $D \to PV$ in the units of $10^{-3}$.
The results from the chromomagnetic penguins 
and quark loops corrections are also listed in the last column.}
 \label{directCP} 
 \vspace{0.2cm}
\scalebox{0.88}{
\centering
\begin{tabular}{cccc}\hline\hline
  ~~~~Modes ~~~~~~~~& ~~~~~~~~Amplitudes ~~~~~~~~ & ~~~~~~~~$A_{CP}$~~~~~~~~ &~~~~ ~~~~$A_{CP}$(+cm,ql) ~~~~ \\\hline
  $D^0$$\to$$\pi^+$$\rho^-$ & $PT$, $PE$       & $-0.04\pm0.01\pm0.00\pm0.00$ & $0.001\pm0.009\pm0.003\pm0.003$ \\

  $D^0$$\to$$\pi^0$$\rho^0$ & $PT$, $PC$, $PE$ & $0.01\pm0.00\pm0.00\pm0.00$ & $-0.06\pm0.00\pm0.00\pm0.00$ \\

  $D^0$$\to$$\pi^0$$\omega$ & $PT$, $PC$, $PE$ & $-0.03\pm0.04\pm0.02\pm0.01$ & $9.08\pm1.43\pm1.07\pm0.62$ \\

  $D^0$$\to$$\pi^0$$\phi$ & $PC$                        & $-0.04\pm0.01\pm0.00\pm0.00$ & $-4.21\pm0.31\pm0.00\pm0.00$ \\

  $D^0$$\to$$\pi^-$$\rho^+$ & $PT$, $PE$         & $0.03\pm0.01\pm0.00\pm0.00$ & $-0.08\pm0.01\pm0.00\pm0.00$ \\

  $D^0$$\to$$K^+$$K^{*-}$ & $PT$, $PE$         & $0.07\pm0.01\pm0.01\pm0.01$ & $0.03\pm0.01\pm0.01\pm0.01$ \\

  $D^0$$\to$$K^0$$\overline{K}^{*0}$    & $PE$           & $-1.07\pm0.16\pm0.00\pm0.01$ & $-1.07\pm0.16\pm0.00\pm0.01$ \\

  $D^0$$\to$$\overline{K}^0$$K^{*0}$   & $PE$            &$-1.07\pm0.16\pm0.00\pm0.01$ & $-1.07\pm0.16\pm0.00\pm0.01$ \\

  $D^0$$\to$$K^-$$K^{*+}$ & $PT$, $PE$      & $-0.02\pm0.01\pm0.00\pm0.00$ & $0.15\pm0.01\pm0.00\pm0.00$ \\

  $D^0$$\to$$\eta$$\rho^0$ & $PT$, $PC$, $PE$,  & $1.02\pm0.13\pm0.09\pm0.04$ & $1.06\pm0.15\pm0.08\pm0.04$ \\

  $D^0$$\to$$\eta$$\omega$ & $PT$, $PC$, $PE$  & $-0.16\pm0.03\pm0.00\pm0.00$ & $1.68\pm0.09\pm0.18\pm0.06$ \\

  $D^0$$\to$$\eta$$\phi$ & $PC$, $PE$               & $-0.04\pm0.00\pm0.01\pm0.01$ & $-1.00\pm0.43\pm0.74\pm0.36$ \\

  $D^0$$\to$$\eta'$$\rho^0$ & $PT$, $PC$, $PE$  & $-0.02\pm0.03\pm0.00\pm0.00$ & $0.29\pm0.06\pm0.04\pm0.01$ \\

  $D^0$$\to$$\eta'$$\omega$ & $PT$, $PC$, $PE$  & $0.29\pm0.23\pm0.02\pm0.01$ & $5.37\pm0.56\pm0.58\pm0.19$ \\

  $D^0$$\to$$\eta'$$\phi$ & $PT$, $PC$, $PE$ & $-0.85\pm0.82\pm0.06\pm0.01$ & $-4.45\pm2.36\pm0.01\pm0.04$ \\

  $D^+$$\to$$\pi^+$$\rho^0$ & $PT$, $PC$         & $-1.65\pm0.12\pm0.46\pm0.16$ & $-1.35\pm0.11\pm0.63\pm0.22$ \\

  $D^+$$\to$$\pi^+$$\omega$ & $PT$, $PC$       & $0.32\pm0.07\pm0.30\pm0.10$ & $-10.32\pm2.60\pm0.58\pm0.43$ \\

  $D^+$$\to$$\pi^+$$\phi$ & $PC$                        & $-0.04\pm0.01\pm0.00\pm0.00$ & $-4.15\pm0.31\pm0.00\pm0.00$ \\

  $D^+$$\to$$\pi^0$$\rho^+$ & $PT$, $PC$         & $-0.15\pm0.04\pm0.06\pm0.02$ & $-0.40\pm0.04\pm0.09\pm0.03$ \\

  $D^+$$\to$$K^+$$\overline{K}^{*0}$ & $PT$       & $0.25\pm0.05\pm0.04\pm0.02$ & $0.21\pm0.05\pm0.03\pm0.02$ \\

  $D^+$$\to$$\overline{K}^0$$K^{*+}$ & $PT$       & $0.08\pm0.02\pm0.01\pm0.00$ & $0.24\pm0.02\pm0.01\pm0.01$ \\

  $D^+$$\to$$\eta$$\rho^+$ & $PT$, $PC$        & $4.00\pm1.31\pm2.76\pm1.06$ & $4.11\pm1.50\pm2.42\pm0.94$ \\

  $D^+$$\to$$\eta'$$\rho^+$ & $PT$, $PC$        & $-0.04\pm0.01\pm0.01\pm0.00$ & $0.28\pm0.03\pm0.03\pm0.01$ \\

  $D_s^+$$\to$$\pi^+$$K^{*0}$ & $PT$              & $-0.22\pm0.04\pm0.03\pm0.02$ & $-0.17\pm0.05\pm0.03\pm0.02$ \\

  $D_s^+$$\to$$\pi^0$$K^{*+}$ & $PT$, $PC$      & $-0.58\pm0.06\pm0.05\pm0.04$ & $-0.52\pm0.06\pm0.05\pm0.04$ \\

  $D_s^+$$\to$$K^+$$\rho^0$ & $PT$, $PC$          & $0.07\pm0.01\pm0.02\pm0.01$ & $-0.08\pm0.01\pm0.02\pm0.01$ \\

  $D_s^+$$\to$$K^+$$\omega$ & $PT$, $PC$          & $-0.13\pm0.02\pm0.02\pm0.01$ & $5.06\pm0.24\pm0.06\pm0.03$ \\

  $D_s^+$$\to$$K^+$$\phi$ & $PT$, $PC$           & $0.28\pm0.04\pm0.13\pm0.04$ & $3.40\pm1.00\pm3.92\pm1.23$ \\

  $D_s^+$$\to$$K^0$$\rho^+$ & $PT$                & $0.10\pm0.01\pm0.01\pm0.01$ &$-0.01\pm0.01\pm0.01\pm0.01$ \\

  $D_s^+$$\to$$\eta$$K^{*+}$ & $PT$, $PC$         & $-0.16\pm0.02\pm0.06\pm0.02$ & $0.03\pm0.05\pm0.06\pm0.02$ \\

  $D_s^+$$\to$$\eta'$$K^{*+}$ & $PT$, $PC$     & $-0.06\pm0.02\pm0.02\pm0.01$ & $-0.15\pm0.02\pm0.05\pm0.02$ \\
  \hline\hline
 \end{tabular}
}
\end{table}
Penguin contributions are highly suppressed and consequently 
cannot interfere effectively with tree amplitudes to generate large  
$\mathit{CP}$ asymmetries. To date, no $\mathit{CP}$ asymmetries in $D \to PV$ decays 
have been definitively measured in experiments.
Nevertheless, as shown in Tab.\ref{directCP}, the $\mathit{CP}$ asymmetries 
in several modes including
$D^0 \to K^0\, \bar K^{*0},\,  \bar K^0\,  K^{*0},\,  
\eta \rho^0, \,  \eta^\prime \phi$, 
$D^+ \to \pi^+ \rho^0, \, \eta \rho^+$, 
and $D_s^+ \to \pi^0 K^{*+}$ are predicted to reach $\mathcal{O}(10^{-3})$,
making them promising candidates for future observation at LHCb, Belle II, or BES III.
Most $\mathit{CP}$ asymmetries remain at $\mathcal{O}(10^{-5})$,
while only a few modes reach $\mathcal{O}(10^{-4})$.
For instance, the decays $D^0 \to \pi^+\rho^-, \, \pi^-\rho^+, \, K^+ K^{*-}$, 
and $K^- K^{*+}$ do not attain the same magnitude as 
their $D \to PP$ counterparts with the same quark-level transitions, 
such as $D^0 \to \pi^+\pi^-$, and $K^+ K^{-}$, which are 
$\mathcal{O}(10^{-4})$~\cite{Li:2012cfa}.
This difference can be attributed to the distinct strong phases 
arising from the $E$ and $PE$ topologies in $D \to PV$ decays
compared to those in $D \to PP$ decays.

The corrections from the chromomagnetic penguins 
and quark loops are substantial, as they can
significantly change the size or even flip the signs
of $\mathit{CP}$ asymmetries.
A notable example is found in the decays
$D^0 \to \pi^0 \phi$ and $D^+ \to \pi^+ \phi$,
where asymmetries involving $C_{PV}$ and $PC_{PV}$ topologies
are enhanced from $\mathcal{O}(10^{-5})$ to $\mathcal{O}(10^{-3})$.
This enhancement occurs because the chromomagnetic penguins
and quark loops corrections to $C_{4,6}$ increase the magnitude  
of $PC_{PV}$ amplitude, thereby strengthening its interference with 
the $C_{PV}$ amplitude.
Consequently, precise measurements of $\mathit{CP}$ asymmetries 
are crucial in the search for new physics, which could manifest through
modifications to the penguin Wilson coefficients similar to those induced 
by the chromomagnetic penguins and quark loopscorrections corrections 
within the Standard Model.


 \section{Conclusion}\label{sec:4}
Driven by extensive new measurements and improved precision 
of charm decays, we present an updated analysis of two-body 
$D \to PV$ decays under the framework of FAT.
The nonfactorizable contributions to all $D \to PV$ decay 
amplitudes are described by only 15 universal parameters 
after factoring out form factors and decay constants, 
capturing flavor SU(3) breaking effects.
In contrast to earlier FAT analyses, we introduce an 
additional nonfactorizable parameter $\chi^{T}$ for the 
$T$ diagram, which we find to be non-negligible 
given cuurent high-precision data even in the 
so-called ``factorizable $T$ diagram. All parameters 
including $\chi^{T}$, the nonfactorizable topological amplitudes,
 $\chi_{V(P)}^C , \, \chi_{q(s)}^E$, $ \chi_{q(s)}^A $,
their associated strong phase $ \phi_{V(P)}^C , \, 
\phi_{q(s)}^E $, $\phi_{q(s)}^A $, the Glauber phase $ S_\pi$, 
and the soft scale $\Lambda_\mathrm{QCD}$, are fitted globally 
from 41 $D \to PV$ branching ratios measurements.
Although the $T$ topology remains dominant, the $C$ topology 
also exhibits substantial nonfactorizable effects.
The parameters $\chi_P^{C}$ and $\chi_V^{C}$ for the 
$C$ amplitude show only minor deviations from earlier 
FAT studies. More pronounced discrepancies are observed 
in the subleading power $E$ and $A$ topologies, which are 
governed by nonfactorizable dynamics. These small values 
of $\chi_{q,s}^{E,A}$ together with their strong phases 
$\phi_{q,s}^{E,A}$ and the Glauber phase $S_\pi$
 highlight the need for more high-precision data to properly 
 constrain these contributions. 
 
Using the updated nonfactorizable parameters, we predict 
$D \to PV$ branching ratios with improved precision, 
yielding results consistent with current experimental data 
and the latest results in topological diagram approach.
We also calculate the direct {\it CP} asymmetries 
arising from interference between tree and penguin diagrams. 
Owing to the substantial updates in the determined strong phases,
our {\it CP} asymmetry predictions differ notably from earlier 
FAT values. Several {\it CP} asymmetries 
are predicted to as large as $\mathcal{O}(10^{-3})$, rendering 
them promising observables for future high-precision experiments. 
Predictions are also provide tor unobserved decay modes, 
particularly those with branching fractions 
in the $10^{-4}-10^{-3}$ range, offering key target for
upcoming experimental studies.

\section*{Acknowledgments}
We are grateful to Qin Qin for useful discussions.
The work is supported by the National Natural Science Foundation of China
under Grants No.12465017 and No.12105148.

\bibliographystyle{bibstyle}
\bibliography{refs}

\end{document}